\documentclass[twocolumn]{pasj00}
\usepackage[varg]{txfonts}
\color{black}

\input{colordvi.tex}

\begin{document}
\Received{$\langle$reception date$\rangle$}
\Accepted{$\langle$accepted date$\rangle$}
\Published{$\langle$publication date$\rangle$}
\SetRunningHead{H. Akamatsu et al.}
{$Suzaku$ X-Ray observations of the accreting NGC~4839 group of galaxies and the radio relic in the Coma cluster}
\title{
$Suzaku$ X-Ray Observations of the Accreting NGC~4839 Group of Galaxies
\\and the Radio Relic in the Coma Cluster
\thanks{
Last update: \today}}
\author{
Hiroki 	{\sc Akamatsu}\altaffilmark{1},
Susumu 	{\sc Inoue}\altaffilmark{2,3},
Takuya {\sc Sato}\altaffilmark{4},
Kyoko 	{\sc Matsusita}\altaffilmark{4},
Yoshitaka {\sc Ishisaki}\altaffilmark{5},
Craig L. {\sc Sarazin}\altaffilmark{6}
}
\altaffiltext{1}{
SRON Netherlands Institute for Space Research, Sorbonnelaan 2, 3584 CA Utrecht, The Netherlands
}
\altaffiltext{2}{
Max-Planck-Institut f\"ur Kernphysik, Saupfercheckweg 1, 69117 Heidelberg, Germany
}
\altaffiltext{3}{
Institute for Cosmic Ray Research, University of Tokyo, 5-1-5 Kashiwanoha, Kashiwa 277-8582, Chiba, Japan
}
\altaffiltext{4}{
   Department of Physics, Tokyo University of Science, 1-3 Kagurazaka, Shinjuku-ku, Tokyo, 162-8601, Japan
   }
\altaffiltext{5}{
  Department of Physics, Tokyo Metropolitan University, 1-1
  Minami-Osawa, Hachioji, Tokyo 192-0397}
\altaffiltext{6}{
Department of Astronomy, University of Virginia, P.O. Box 400325, Charlottesville, VA 22904-4325, USA}
\email{H.Akamatsu@sron.nl}
\maketitle
\begin{abstract}
Based on Suzaku X-ray observations, we study the hot gas around the NGC4839 group of galaxies and the radio relic 
in the outskirts of the Coma cluster.  
We find a gradual decline in the gas temperature from 5 keV around NGC4839 to 3.6 keV at the radio relic, 
across which there is a further, steeper drop down to 1.5 keV. 
This drop as well as the observed surface brightness profile are consistent with a shock 
with Mach number ${\cal M}=2.2 \pm 0.5$ and velocity $v_s = (1410\pm110)\rm ~km~ s^{-1}$. 
A lower limit of
 $B > 0.33 ~\mu$G 
 is derived on the magnetic field strength around the relic 
from upper limits to inverse Compton X-ray emission. 
Although this suggests that the non-thermal electrons responsible for the relic are generated 
by diffusive shock acceleration (DSA), the relation between the measured Mach number and 
the electron spectrum inferred from radio observations are inconsistent with that expected from the simplest, 
test-particle theory of DSA. Nevertheless, DSA is still viable if it is initiated by the injection of a pre-existing population of 
non-thermal electrons. 
Combined with previous measurements, the temperature profile of Coma 
in the southwest direction is shallower outside NGC4839 and also slightly shallower in the outermost region. 
The metal abundance around NGC4839 is confirmed to be higher than in its vicinity, implying a significant peak in 
the abundance profile that decreases to ~0.2 solar toward the outskirts. 
We interpret these facts as due to ram pressure stripping of metal-enriched gas from NGC4839 as it falls into Coma. 
The relic shock may result from the combined interaction of pre-existing intracluster gas, gas associated with NGC 4839, 
and cooler gas flowing in from the large-scale structure filament in the southwest.
\end{abstract}

\begin{figure*}[ht]
\begin{tabular}{cc}
\begin{minipage}{0.5\hsize}
\begin{center}
\includegraphics[width=1.\hsize]{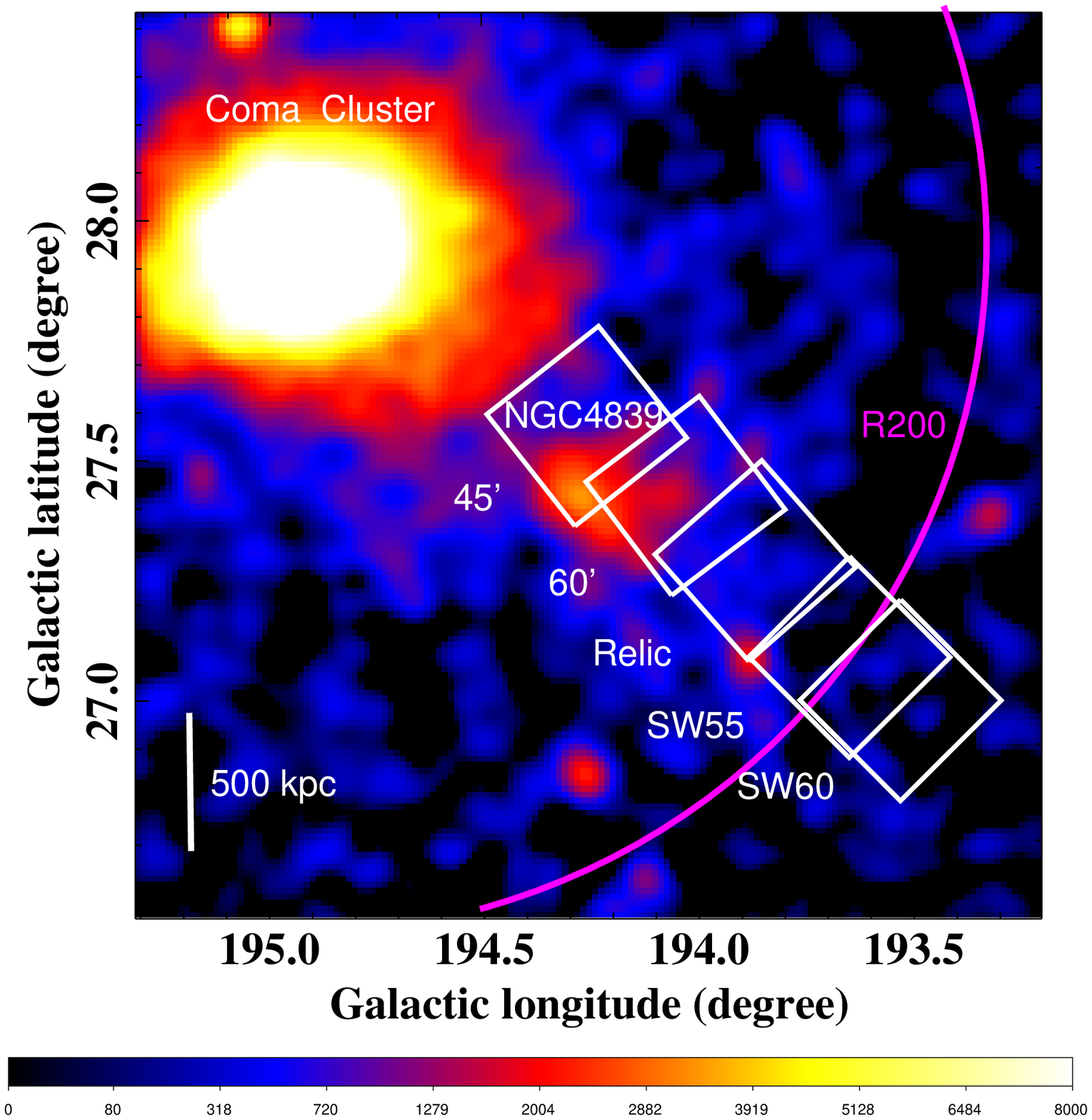}
\end{center}
\end{minipage}
\begin{minipage}{.5\hsize}
\begin{center}
\includegraphics[height=1.\hsize]{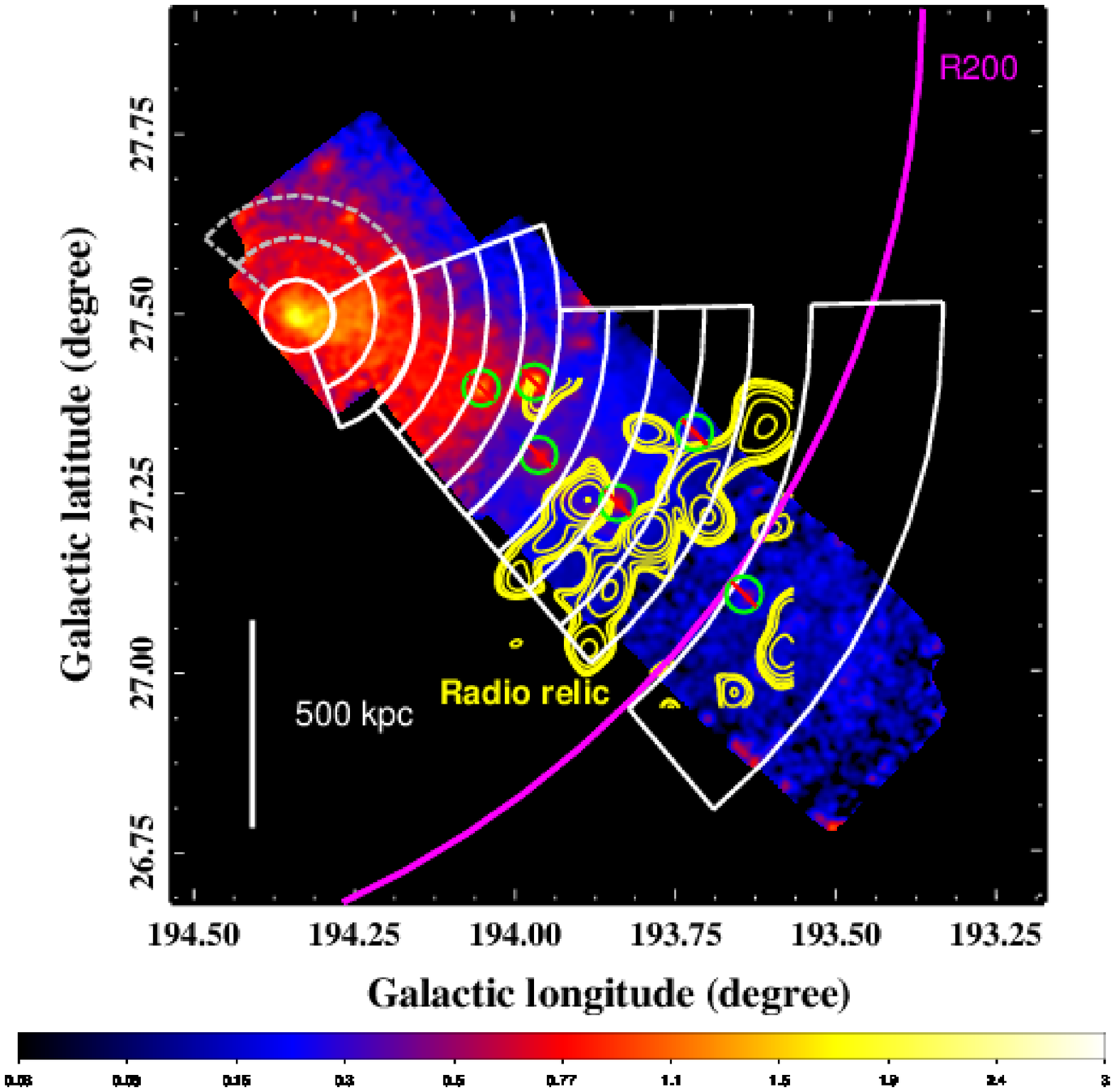}
\end{center}
\end{minipage}
\end{tabular}
\caption{\label{fig:image}
Left: Background subtracted ROSAT All Sky Survey image in the 0.1-2.4 keV band
of the southwestern regions of the Coma cluster.
The color scale unit is cts Ms$^{-1} \rm pixel^{-2}$ (1 pixel=$\timeform{45''}\times \timeform{45''}$). 
The white boxes show the FOVs of the Suzaku XIS observations we discuss,
while the magenta circle shows Coma's virial radius $r_{200}$,
corresponding to $\timeform{83.9'}$ from the cluster center.
Right:
NXB subtracted Suzaku FI+BI mosaic image in 0.5-8.0 keV band, smoothed with a
  2-dimensional gaussian with $\sigma =16$ pixel =$\timeform{17''}$ . 
  The image is corrected for exposure time but not for vignetting. 
The color scale unit is cts Ms$^{-1} \rm pixel^{-2}$ (1 pixel=$\timeform{1.04''}\times\timeform{1.04''}$).  
  The white and gray dashed annuli show the regions used for the spectral analysis.  
   Small green circles show point sources. 
 The NVSS 1.4 GHz image of the radio relic is shown in yellow contours.
}
\end{figure*}

\begin{table*}[ht]
\begin{center}
\caption{\label{tab:obslog}
Observation log and exposure time after data screening without and with COR-cut (COR2 $>$ 6 GV) }
\begin{tabular}{lcccc}  \hline   \
Name (Obs I.D) &(R.A., DEC) & Observation Date & Exp. Time$^{\dagger}$ & Exp. Time$^{\ddagger}$ \\ \hline  
$45'$ (802047010) & (194.26, 27.58) & 2007-12-02T10:08:56 & 28.0 & 22.5  \\
$60'$ (802048010) & (194.03, 27.44) & 2007-12-04T04:47:49 & 32.9 & 27.6  \\
Radio relic (803051010) & (193.86, 27.30) & 2008-12-23T21:06:23 & 172.3 & 145.9 \\ 
SW55 (806048010) & (193.64, 27.07) & 2011-06-23T20:47:37 & 15.2 & 14.6  \\
SW60 (806049010) & (193.53, 26.98) & 2011-06-24T07:17:49 & 12.7 & 8.8  \\
COMABKG (802083010) & (198.75, 31.63) & 2007-06-21T00:14:27 & 28.2 & 24.1 \\
\hline 
$\dagger$: COR2  $>$ 0 GV &\multicolumn{2}{l}{ $^{\ddagger}$: COR2 $>$ 6 GV}\\\
\end{tabular}
\end{center}
\end{table*}

\section{Introduction}
\label{sec:intro}
According to modern theories of large scale structure formation,
clusters of galaxies are built up hierarchically through successive mergers and accretion 
of smaller subsystems \citep{voit05,kravtsov12}.
In particular, major mergers involving multiple subclusters of comparable masses
(hereafter simply ``mergers'' unless otherwise indicated)
are dramatic events that strongly affect the intracluster medium (ICM) and
can convert as much as $10^{64}$ erg of kinetic energy into thermal energy through shocks and turbulence \citep{markevitch07}.
Shocks not only heat the gas but can also produce high-energy, nonthermal particles
through the mechanism of diffusive shock acceleration (DSA) \citep{blandford87}
and provide a key observational tool in the study of cluster growth.
X-ray observations have revealed direct evidence of shock fronts
through discontinuities in the surface brightness and temperature
in the ICM of  merging clusters,
for example the so-called ``Bullet'' cluster 1E 0657-56 \citep{markevich02},
and more recently in Abell~2146 \citep{russell10} and Abell~754 \citep{macario11}.
Such shocks discovered so far in X-rays are associated with clusters undergoing relatively major mergers \citep{markevitch10}.
Large-scale shocks driven by the accretion of diffuse gas that is not bound in virialized systems
are also theoretically expected and generally seen in numerical simulations \citep{miniati00,ryu03},
but have yet to be observationally confirmed in X-rays.

The Coma cluster is one of the best-studied clusters of galaxies
because of its low redshift ($z=0.023$) and its high richness \citep{biviano98}.
In view of its inhomogeneous distribution of galaxies and disturbed X-ray surface brightness,
it is acknowledged to be a merging system \citep{hughe88, hughe93}.
The ROSAT X-ray image clearly shows some substructures \citep{briel91}.
\citet{watanabe99} presented temperature maps based on ASCA data
that revealed the ICM distribution to be non-isothermal.
\citet{neumann01} reported the structure of Coma on large scales based on XMM-Newton mosaic observations,
which indicate that it is now undergoing at least one subcluster-cluster merger.
In particular, the group of galaxies associated with the elliptical galaxy NGC~4839 is a prominent substructure
located southwest of the cluster center, and may be falling into Coma \citep{adami05}.
\citet{neumann01} also found indications of a bow shock and ram pressure stripping around NGC~4839.

Merging clusters including Coma also often exhibit radio halos and/or radio relics,
which are spatially-extended, synchrotron emission from non-thermal populations of electrons
that are somehow accelerated over scales of up to $\sim$Mpc \citep{ferrari08,feretti12}.
Radio halos occur in the central regions with a fairly symmetric morphology,
and are often considered to be a result of stochastic acceleration by merger-induced turbulence (e.g., \cite{brunetti11}; 
see however \cite{dennison80}).
By contrast, radio relics are found near the cluster periphery with a tangentially-elongated morphology, 
and may be due to DSA in accretion or merger shocks 
(e.g. \cite{ensslin98,miniati01,brueggen12, kang12,skillman13, pinzke13}).

The Coma radio relic is located near its virial radius about 2 Mpc in projection
from the cluster center, beyond NGC4839 but in the same southwest direction.
\citet{feretti06} investigated the relic region with XMM-Newton observations.  
Although they detected thermal emission around the radio relic, they did not find
any temperature enhancement or evidence of a shock in regions northeast of the relic towards the cluster center.
Thus they proposed that the relativistic electrons are accelerated by turbulence generated by the infall of the NGC~4839 group,
rather than by shock acceleration.
\citet{brown11} discussed the possible connection with ongoing accretion based on 
their radio and optical observations.
Recently, $Planck$ found signatures of shock fronts through measurements of the Sunyaev-Zeldovich (SZ) effect
in two regions about half a degree to the west
and to the south-east of the cluster center, which correspond to the edge of the radio halo~\citep{planck12}. 
They derive  Mach numbers ${\cal M}_{\rm W}= 1.95^{+0.45}_{-0.02}$ and 
${\cal M}_{\rm SE} = 2.03^{+0.14}_{-0.04}$ for the shocks in the west and southeast, respectively.
Gamma-ray observations searching for evidence of high-energy particles have also been conducted for Coma, 
but have not been successful so far 
(\cite{aharonian09, ackermann10, arlen12}; see however \cite{keshet12}).

Possible associations between shocks and radio relics
have also been found in a number of other merging clusters.
\citet{vanweeren10} reported the 
discovery of a double radio relic with an unprecedentedly narrow morphology in a merging cluster.
The spatially resolved radio spectral index is $\alpha\sim-0.6$
along the sharp outer edges of the relics,
which presumably reflects the spectrum of the emitting electrons
that have just been accelerated and are yet to be affected by radiative losses.
Under the assumption that these electrons are generated according to
the simplest, test-particle theory of DSA,
the spectral index can be related to the Mach number of the shock, which in this case would be ${\cal M}\sim4.5$.
Subsequent, deep X-ray follow-up observations revealed the existence of a shock front at the radio relic~\citep{ogrean12, akamatsu12}.
The cluster Abell~3667 has a pair of radio relics, of which the northwestern one
is the brightest diffuse cluster radio source in the sky \citep{roettgering97}.
\citet{finoguenov10} found evidence of a merger-induced shock at the outer edge of this relic,
and showed that it was energetically capable of powering the radio relic.
Thus, it appears that radio relics can be good tracers of merger shocks in clusters.

In this paper, we discuss X-ray observations focusing on the regions of interaction between Coma and
NGC~4839 and the radio relic located near the outskirts of the Coma cluster.
We assume cosmological parameters $H_0 = 70$ km s$^{-1}$ Mpc$^{-1}$, $\Omega_{\rm M}=0.27$ and
$\Omega_\Lambda = 0.73$, which implies 28.9 kpc per arcminute at Coma's redshift of $z = 0.023$.  
The virial radius is approximately $r_{200} = 2.77 h_{70}^{-1} (\langle T\rangle /10 \, {\rm keV})^{1/2}/E(z)\ {\rm Mpc} $,
where $E(z)=[\Omega_{\rm M}(1+z)^{3}+1-\Omega_{\rm M}]^{1/2}$ \citep{henry09}.  
For our cosmology  and average temperature $k \langle T \rangle = 7.8$ keV as used in \citet{matsushita11},
$r_{200}=$ 2.43 Mpc, corresponding to $\timeform{83.9'}$ from the cluster center.
As our fiducial reference for the solar photospheric abundances 
denoted by $Z_\odot$, 
we adopt \citet{lodders03}.
Unless otherwise stated, the errors correspond to 68\% confidence for a single parameter.

\section{Observations and Data Reduction}\label{sec:data}
As shown in Fig.\ \ref{fig:image}, Suzaku carried out five observations
with different pointings of the southwest region of the Coma cluster in December 2007, 2008 and July 2011.
They covered the regions from around
NGC4839 to beyond the radio relic.
We refer to these pointings as $45'$, $60'$,  Radio Relic, SW55 and SW60.
The observation log is given in Table~\ref{tab:obslog}.  
All the observations were performed with either normal $5\times5$ or $3\times3$ clocking modes.  
The virial radius $r_{200}$ of the Coma cluster
($2.43$ Mpc corresponding to $\timeform{83.9'}$)
is indicated by the magenta circles in Fig.\ \ref{fig:image}.

\begin{figure}[h]
\begin{center}
 \includegraphics[height=1.\hsize,angle=-90]{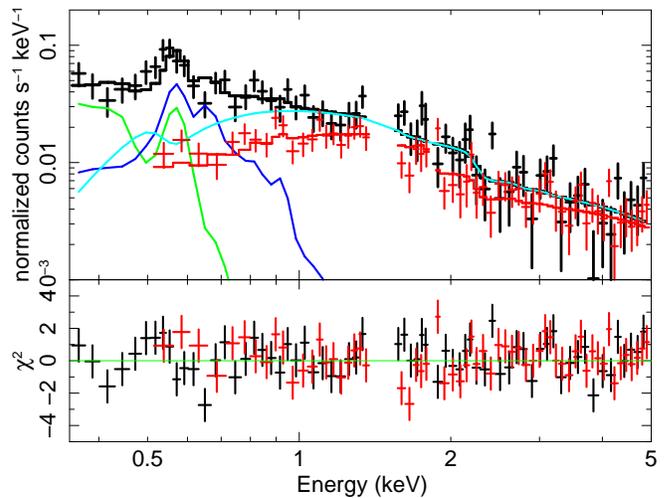}
\end{center}
\caption{The spectrum of the COMA BGK field used for the background
  estimation, after subtraction of the NXB and point sources.
  The XIS BI (black) and FI (red) spectra are fitted with CXB $+$ Galactic
  components (LHB, MWH) [{\it apec+wabs(apec+powerlaw)}].  
The CXB spectrum is shown with a cyan curve, and the LHB and 
MWH components are indicated by green and blue curves, respectively. }
\label{fig:bgd}
\end{figure}

The XIS instrument consists of 4 CCD chips: one back-illuminated (BI: XIS1) and 
three front-illuminated (FI: XIS0, XIS2, XIS3) detectors.  
We used HEAsoft version 6.12 and CALDB 2012-02-10 for all of the Suzaku data analysis presented here.  
We performed event screening with the cosmic-ray cut-off rigidity COR2 $> 6$ GV 
to increase the signal to noise ratio.
We also limited  the Earth rim ELEVATION to $> 10^{\circ}$ to avoid contamination from scattered
solar X-rays from the dayside Earth limb.
For the SW55 and SW60 data, we applied additional processing for the XIS1 detector to reduce the NXB level, which
increased after the change in the amount of charge injection in June 2010.
The detailed processing procedures are the same as those described in $XIS~analysis~topics$\footnote{http://www.astro.isas.jaxa.jp/suzaku/analysis/xis/xis1\_ci\_6\_nxb/}.

\begin{table*}[th]
\caption{ Best-fit background parameters}
\small
\begin{center}
\begin{tabular}{lcccccccccccc}
\hline 
                          & NOMINAL & CXBMIN - 3\% of NXB&CXBMAX + 3\% of NXB&CONTAM-10\% &CONTAM+10\% \\ \hline
 LHB& \\
 $kT$ (keV) &      0.08 (fix)   &0.08 (fix)   &0.08 (fix)   &0.08 (fix)   &0.08 (fix)   &\\
$norm^{\ast}$  $(\times 10^{-3}$)&    $ {9.7}\pm3.2 $  &  11.8	&7.2 &6.9 &14.9\\ \hline
 MWH &                 \\
 $kT$ (keV) &    $  {0.18}\pm0.02  $ & 0.19   &0.14 &0.15 &0.19\\  
$norm^{\ast}$ ($\times 10^{-4}$)&   $  {9.8}\pm2.7 $  &  9.7    &  17.7&15.1 &9.9\\ \hline
CXB ($\Gamma=1.41$ fix) & & \\
$norm$$^\dagger$	($\times 10^{-4}$)& $8.7\pm0.4$	&7.8 (fix)	& 9.5 (fix) &8.7 (fix) &8.7 (fix)\\ \hline
$\chi^{2}$/d.o.f    &  63 / 82& 66 / 83     & 73 / 83    & 63 / 83&63 / 83 \\ \hline
\multicolumn{6}{l}{
$\ast$: Normalization of the apec component scaled by a factor 400$\pi$.}\\
\multicolumn{6}{l}{
Norm=$\rm 400\pi$$\int n_{e}n_{H} dV/(4\pi(1+z^2)D_{A}^2)\times 10 ^{-14} \rm ~cm^{-5}~arcmin^{-2}$, where $D_A$ is the angular diameter distance to the source.}\\
\multicolumn{6}{l}{
$\dagger$: The
CXB intensity normalization in \cite{kushino02} is 9.6$~\times10^{-4}$ for $\Gamma=1.41$ }\\
\multicolumn{6}{l}{
in units of photons keV$^{-1}~\rm cm^{-2}~s^{-1}$ at 1 keV.}
\end{tabular}
\label{tab:bgd}
\end{center}
\end{table*}

We extracted pulse-height spectra in 12 annular regions whose boundaries are given
 in Table~\ref{tab:bestfit}  with the center at ($\timeform{12h57m24s}, \timeform{27D29'54"}$), 
which is the X-ray peak associated with NGC4839~\citep{neumann01}.  
For the~\timeform{43'}$-~\timeform{54'}$ annulus, we combined the SW55 and SW60 data to increase the signal to noise ratio.
There is a missing annulus for \timeform{38'}$-~\timeform{43'}$.
This region was only covered by the SW55 observation, and the shorter exposure and lower surface brightness led to too few counts for spectral fitting.
In all annuli, the positions of the calibration sources were masked out using the {\it calmask} calibration database (CALDB) file.

\begin{table}[bht]
\begin{center}
\caption{\label{tab:bestfit}Best-fit spectral parameters of the ICM}
\begin{tabular}{cccccccccccc}\hline
Region$^{\ast}$ &k$T$ (keV) &$Z$ ($Z_{\odot}$)&norm$^\dagger$ & $\chi^{2}$/d.o.f \\ \hline
$\timeform{0'}-\timeform{3'} $& $ {4.10}^{+0.17}_{-0.13} $ & $  {0.64}^{+0.10}_{-0.10}  $  &  $
{63.3}^{+1.9}_{-2.0}  $  & 181 / 154  \\
$\timeform{3.0'}-\timeform{5.5'} $& $ {5.20}^{+0.23}_{-0.29} $ & $  {0.58}^{+0.10}_{-0.13}  $  &  $
{49.1}^{+1.7}_{-1.6}  $  & 109 / 129  \\
$\timeform{5.5'}-\timeform{10'} $& $ {4.66}^{+0.23}_{-0.30} $ & $  {0.42}^{+0.11}_{-0.14}  $  &  $
{42.3}^{+1.7}_{-1.3}  $  & 110 / 125  \\

$\timeform{10'}-\timeform{13'} $&   $ {3.48}^{+0.16}_{-0.22} $ & $  {0.35}^{+0.08}_{-0.10}  $  &  $  {38.3}^{+1.6}_{-1.7}  $  & 142 / 137  \\
$\timeform{13'}-\timeform{16'} $& $ {3.44}^{+0.13}_{-0.16} $ & $  {0.31}^{+0.06}_{-0.09}  $  &  $  {31.9}^{+1.7}_{-0.7}  $  & 197 / 170  \\

$\timeform{16'}-\timeform{19'} $& $  {3.64}^{+0.17}_{-0.16}  $  &  $  {0.36}^{+0.09}_{-0.08}$  & $ {23.7}^{+0.9}_{-1.0}$  & 183 / 175\\
$\timeform{19'}-\timeform{22'}$& $  {3.47}^{+0.27}_{-0.25}  $  &  $  {0.29}^{+0.11}_{-0.12}$  & $ {17.2}^{+1.3}_{-0.7}$  & 114 / 110\\

$\timeform{22'}-\timeform{26'} $& $  {3.64}^{+0.13}_{-0.12}  $  &  $  {0.28}^{+0.05}_{-0.05}$  & $ {14.2}^{+0.4}_{-0.3}$  & 402 / 360\\
$\timeform{26'}-\timeform{30'} $& $  {3.49}^{+0.16}_{-0.14}  $  &  $  {0.21}^{+0.06}_{-0.05}$  & $ {9.4}^{+0.3}_{-0.3}$  & 395 / 360\\
$\timeform{30'}-\timeform{34'} $& $  {3.57}^{+0.21}_{-0.20}  $  &  $  {0.17}^{+0.07}_{-0.07}$  & $ {7.9}^{+0.3}_{-0.3}$  & 361 / 360\\
$\timeform{34'}-\timeform{38'} $& $  {2.67}^{+0.51}_{-0.22}  $  &  $  {0.17}^{+0.12}_{-0.09}$  & $ {5.5}^{+0.6}_{-0.4}$  & 208 / 174\\
$\timeform{43'}-\timeform{54'} $& ${1.54}^{+0.45}_{-0.39}$ 	&0.15 (fix)	& ${2.3}^{+0.5}_{-0.3}$		& 118 / 147	& \\ 
\hline
\multicolumn{6}{c}{
North Direction (Gray dotted annuli in Fig.~\ref{fig:image}) 
}\\ \hline
$\timeform{3.0'}-\timeform{5.5'} $&$ {6.95}^{+0.75}_{-0.47} $ & $  {0.13}^{+0.13}_{-0.13}  $  &  $
{30.2}^{+1.4}_{-1.0}  $  & 134 / 112  \\
$\timeform{5.5'}-\timeform{10'} $& $ {5.74}^{+0.40}_{-0.51} $ & $  {0.24}^{+0.12}_{-0.12}  $  &  $
{28.2}^{+1.0}_{-1.1}  $  & 119 / 112  \\

\hline
\multicolumn{6}{l}{\footnotesize
$\ast$: Radius from NGC4839, 
}\\
\multicolumn{6}{l}{\footnotesize
$\dagger$:Norm=$\int n_{e}n_{H} dV/(4\pi(1+z^2)D_{A}^2)\times 10 ^{-20} \rm ~cm^{-5}~arcmin^{-2}$, }\\
\multicolumn{6}{l}{\footnotesize
where $D_A$ is the angular diameter distance to the source.}\\
\end{tabular}
\end{center}
\end{table}

\section{Spectral Analysis and Results}\label{sec:spec}
\subsection{Method of Spectral Fits}\label{sec:spec_model}
The observed spectrum was assumed to consist of optically thin thermal plasma emission from the ICM, 
that from the Local Hot Bubble (LHB) and the Milky Way Halo (MWH) as the Galactic foreground components, cosmic
X-ray background (CXB), and non-X-ray background (NXB).  
The NXB component was estimated from the dark Earth database using the ${\it xisnxbgen}$ program in FTOOLS \citep{tawa08},
and was subtracted from the data before the spectral fit.  
To adjust for the long-term variation of the XIS background due to radiation damage, we accumulated the NXB data
for the period between 150 days before and 150 days after each observation.

For the spectral fits, we used XSPEC ver12.7.0.  
To generate the ARF for the \timeform{45'} region and for the \timeform{10'}-\timeform{13'} and 
\timeform{13'}--\timeform{16'} annular regions, 
we used the $xissimarfgen$ program in FTOOLS~\citep{ishisaki07} and 
the ROSAT PSPC image for the simulated surface brightness profile (OBS ID:US800009P.N1). 
For the \timeform{16'}--\timeform{19'} and -\timeform{19'}--\timeform{22'} annular regions and 
the relic, SW55 and SW60 regions,  we generated spatially uniform ARFs 
over a circular region of $\timeform{20'}$  radius for the background and ICM emission.
We carried out spectral fits to the pulse-height spectrum in each annulus separately.  
We analyzed the spectra in the 0.35--7~keV range for the BI detectors and 0.5--8~keV for the FI detectors.
Because of possible contamination by solar wind charge exchange, 
we only analyzed the spectra above 0.9 keV for all detectors in the
SW55 region (\timeform{43'}$-~\timeform{54'}$ annulus).
None of the other spectra were significantly affected by solar wind charge exchange.
In addition, above 5 keV, the CXB dominates the observed spectra of the \timeform{43'}$-~\timeform{54'}$ annulus.
For this region, we only use the spectrum below 5 keV to increase the signal to noise ratio.
As shown by the green circles in Figure~\ref{fig:image}, we found 6 point-like sources and masked out the pixels
within a 1 arcmin radius around these sources.

\begin{figure*}[ht]
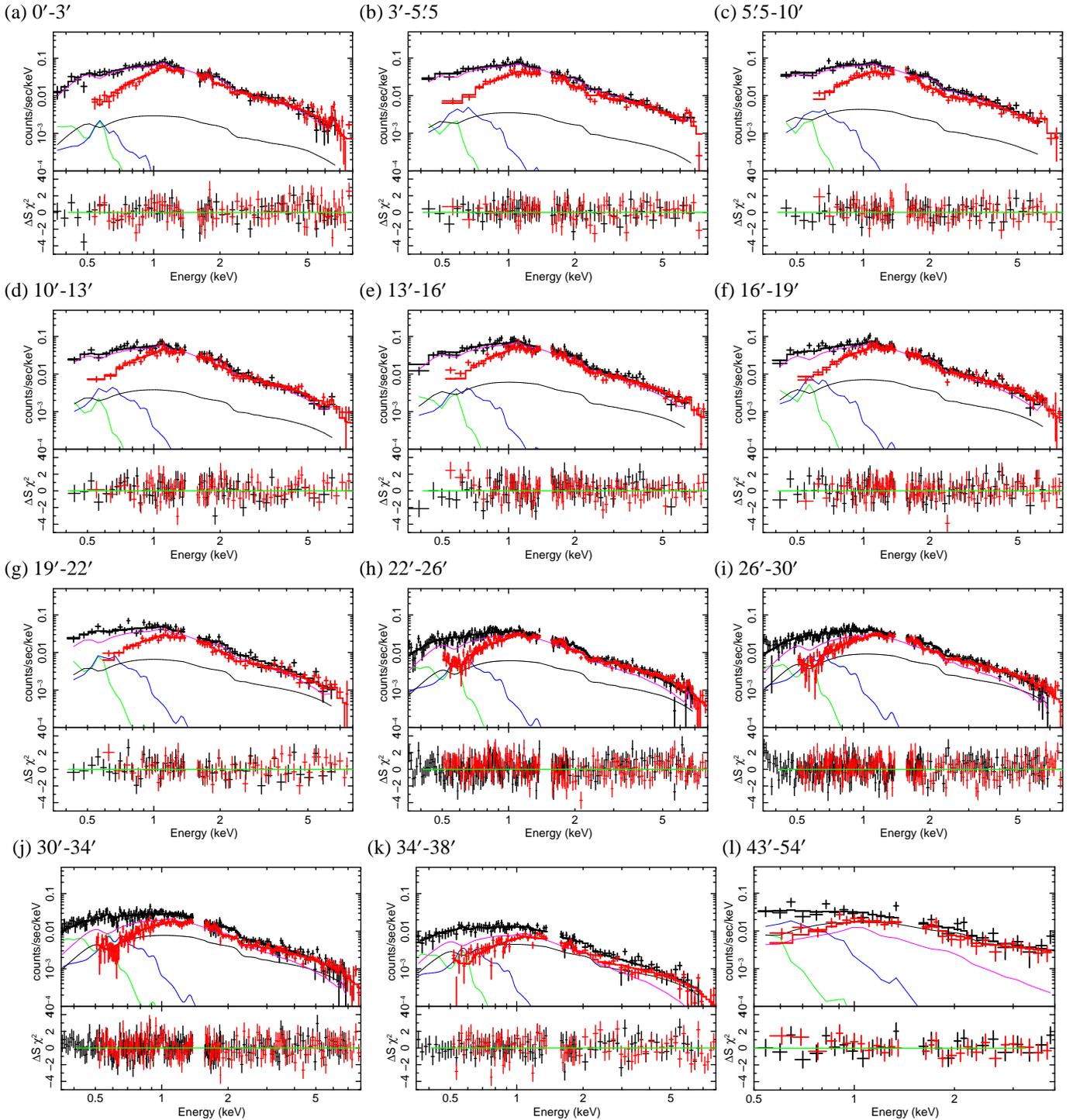

\begin{tabular}{cc}
\begin{minipage}{0.333\hsize}
(a) \timeform{0'}-\timeform{3'}
\\[-0.8cm]
\begin{center}
\includegraphics[angle=-90,width=1.\hsize]{fig/ICM-check-45-1.ps}
\end{center}
\end{minipage}
\begin{minipage}{0.3333\hsize}
(b) \timeform{3'}-\timeform{5.5'}
\\[-0.8cm]
\begin{center}
\includegraphics[angle=-90,width=1.\hsize]{fig/ICM-check-45-2.ps}
\end{center}
\end{minipage}
\begin{minipage}{0.3333\hsize}
(c) \timeform{5.5'}-\timeform{10'}
\\[-0.8cm]
\begin{center}
\includegraphics[angle=-90,width=1.\hsize]{fig/ICM-check-45-3.ps}
\end{center}
\end{minipage}\\
\begin{minipage}{0.3333\hsize}
(d) \timeform{10'}-\timeform{13'}
\\[-0.8cm]
\begin{center}
\includegraphics[angle=-90,width=1.\hsize]{fig/ICM-check-60-4.ps}
\end{center}
\end{minipage}
\begin{minipage}{0.3333\hsize}
(e) \timeform{13'}-\timeform{16'}
\\[-0.8cm]
\begin{center}
\includegraphics[angle=-90,width=1.\hsize]{fig/ICM-check-60-5.ps}
\end{center}
\end{minipage}
\begin{minipage}{0.33333\hsize}
(f) \timeform{16'}-\timeform{19'}
\\[-0.8cm]
\begin{center}
\includegraphics[angle=-90,width=1.\hsize]{fig/ICM-check-60-6.ps}
\end{center}
\end{minipage}\\
\begin{minipage}{0.3333\hsize}
(g) \timeform{19'}-\timeform{22'}
\\[-0.8cm]
\begin{center}
\includegraphics[angle=-90,width=1.\hsize]{fig/ICM-check-60-7.ps}
\end{center}
\end{minipage}
\begin{minipage}{0.3333\hsize}
(h) \timeform{22'}-\timeform{26'}
\\[-0.8cm]
\begin{center}
\includegraphics[angle=-90,width=1.\hsize]{fig/ICM-check-relic-81.ps}
\end{center}
\end{minipage}
\begin{minipage}{0.33333\hsize}
(i) \timeform{26'}-\timeform{30'}
\\[-0.8cm]
\begin{center}
\includegraphics[angle=-90,width=1.\hsize]{fig/ICM-check-relic-82.ps}
\end{center}
\end{minipage}\\
\begin{minipage}{0.33333\hsize}
(j) \timeform{30'}-\timeform{34'}
\\[-0.8cm]
\begin{center}
\includegraphics[angle=-90,width=1.\hsize]{fig/ICM-check-relic-91.ps}
\end{center}
\end{minipage}
\begin{minipage}{0.33333\hsize}
(k) \timeform{34'}-\timeform{38'}
\\[-0.8cm]
\begin{center}
\includegraphics[angle=-90,width=1.\hsize]{fig/ICM-check-relic-10.ps}
\end{center}
\end{minipage}
\begin{minipage}{0.318\hsize}
(l) \timeform{43'}-\timeform{54'}
\\[-0.8cm]
\begin{center}
\includegraphics[angle=-90,width=1.\hsize]{fig/ICM-check-2obs.ps}
\end{center}
\end{minipage}
\end{tabular}
\caption{
NXB subtracted spectra in each annular region.
The XIS BI (Black) and FI (Red) spectra are fitted with the ICM model (Magenta: {\it wabs $\times$ apec}),
along with the sum of the CXB and the Galactic emission [{\it apec + wabs(apec + powerlaw)}].
The CXB component is shown with a black curve, and the LHB and MWH emissions are indicated
by green and blue curves, respectively. For the \timeform{43'}-\timeform{54'} annulus, we show the SW60 spectrum.
}
\label{fig:fit}
\end{figure*}

In order to estimate the CXB and Galactic foreground components, 
we selected the offset Suzaku observation of COMA BGK (Observation ID = 802083010), 
which was centered at 6 degrees angular distance from NGC~4839.
Although the background components were estimated in \citet{takei08} and \citet{sato11},
we re-analyzed the same data to determine the systematic errors.
We modeled the sky backgrounds as a combination of
the Local Hot Bubble (LHB $\sim$ 0.08 keV), the Milky Way Halo (MWH $\sim$ 0.3 keV),
and the cosmic X-ray background (CXB), using the spectral model: ${\it apec+wabs(apec+powerlaw)}$.
In the fits, we fixed the temperature of the LHB component to 0.08 keV.
The redshift and metal abundance of both the {\it apec} components were fixed at 0 and solar, respectively.
The resultant spectrum and parameters are shown in Figure~\ref{fig:bgd} and Table~\ref{tab:bgd}.
The temperature of the MWH is $0.18\pm0.02$ keV. 
The temperature and intensity are consistent with the typical Galactic emission and 
previous measurements~\citep{yoshino09, takei08, sato12}.  
For the estimate of the systematic error of the background spectrum, we adopted the amplitude of the CXB fluctuations as 10\%,
the uncertainty of the NXB intensity as 3\%  and the uncertainty in the contamination
on the optical blocking filter to be $\pm 10\%$, respectively.  
Using this background model, we also examined the effect of systematic errors on our spectral parameters (Sec.~\ref{sec:sys}).

The normalization of the LHB component, the normalization and
temperature of the  MWH component, and the normalization of the power-law model for the CXB component 
were allowed to vary within the range of the errors in the background estimate~(Table~\ref{tab:bgd}). 
We employed a single temperature thermal model ($wabs\times apec$) for the ICM emission of the Coma cluster.
In the radio relic region, in addition to the above model, 
we employed a two-temperature
thermal model to account for the possible effects of PSF blurring 
and projection (see sec.~\ref{sec:result}).
The interstellar absorption was fixed using the 21 cm measurement of the hydrogen column density,
$N_{\rm H}=1.0\times10^{20}~\rm cm^{-2}$~\citep{dickey90}.
The redshift of the ICM component was fixed at 0.023~\citep{biviano98}.
In the inner region ($r<38'$ from NGC4839), the temperature, abundance, and normalization of
the single-temperature model were free parameters.
In the outer regions ($r>\timeform{38'}$), we fixed the ICM metal abundance to 0.15 solar.
In the simultaneous fit of the BI and FI data, only the normalizations were allowed to differ, 
although we found that the derived normalizations were consistent within 15\%. 
In general, each data set was fitted well with the above mentioned model.

\begin{figure*}[thb]
\begin{tabular}{cc}
\begin{minipage}{0.5\hsize}
(a) Temperature
\\[-0.8cm]
\begin{center}
\includegraphics[width=1.\hsize,angle=-90]{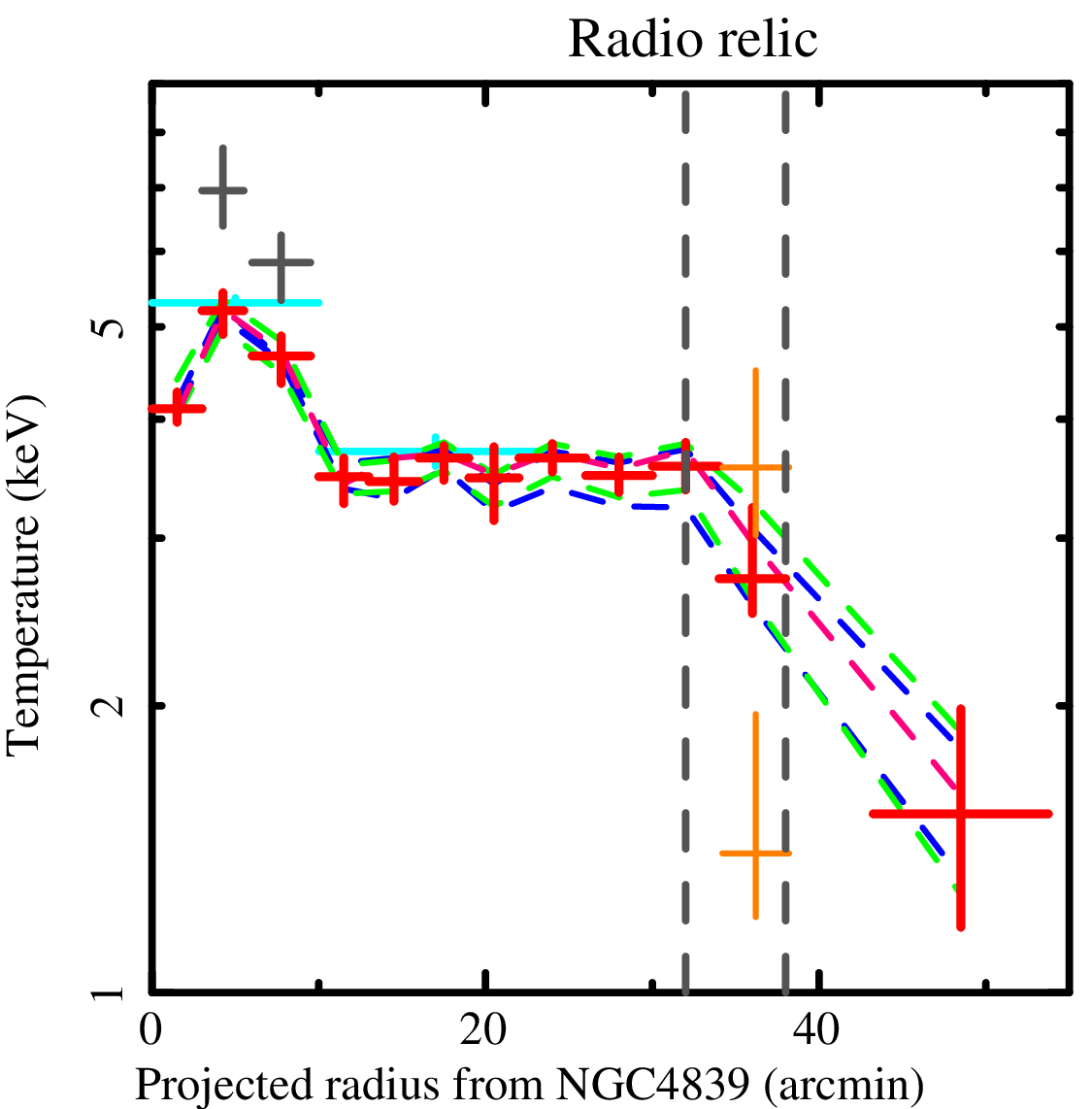}
\end{center}
\end{minipage}
\begin{minipage}{0.5\hsize}
(b) Metal abundance
\\[-0.8cm]
\begin{center}
\includegraphics[width=.9\hsize,angle=-90]{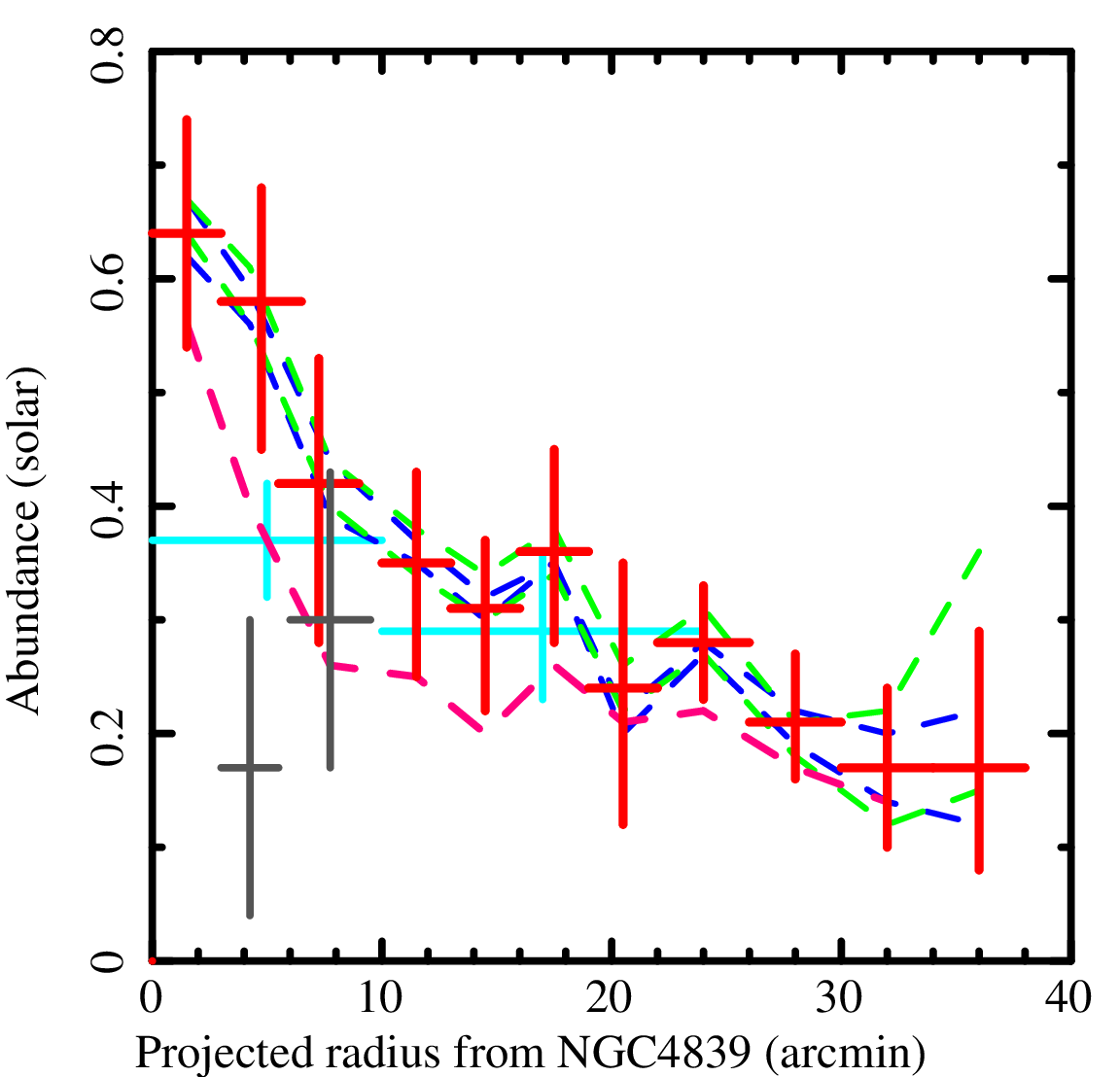}
\end{center}
\end{minipage}\\
\end{tabular}
\caption{ 
Radial profiles of the (a) ICM temperature, and (b) metal abundance, centered on NGC4839.
The Suzaku best-fit values with statistical errors are shown with red crosses.  
Gray crosses show the results for the northern regions (the gray dotted annuli in Figure~\ref{fig:image}).
The gray dashed vertical lines show the approximate radial boundaries of the radio relic (1253+275). 
The cyan crosses show the previous measurements by ~\citet{sato11},
with converted 1$\sigma$ errors.
The orange crosses show the the result for the two temperature model.
The range of uncertainties due to the combined 3\% variation of the 
NXB level and the maximum/minimum fluctuation in the CXB is shown by two blue dashed curves.
In a similar way, the range of uncertainties due to
the contamination on the blocking filter is shown by green dashed curves.
Magenta dashed lines show changes due to different abundance tables.  
 }
\label{fig:radial}
\end{figure*}

\begin{table*}[bht]
\caption{Estimation of the CXB fluctuation.}
\centering
\footnotesize
\begin{tabular}{ccccccccccccccccccc}\hline
Region$^\ast$ &
\timeform{0'}-\timeform{3'}&
 \timeform{3.0'}-\timeform{5.5'}&
 \timeform{5.5'}-\timeform{10'}&
 \timeform{10'}-\timeform{13'}&
 \timeform{13'}-\timeform{16'}&
 \timeform{16'}-\timeform{19'}&
 \timeform{19'}-\timeform{22'}&
 \timeform{22'}-\timeform{26'}&
 \timeform{26'}-\timeform{30'}&
 \timeform{30'}-\timeform{34'}&
 \timeform{34'}-\timeform{38'}&
 \timeform{43'}-\timeform{54'}
\\ \hline
Area$^\dagger$	 &
28.3	&
60.9	&
95.9	&
55.7 &
73.0 &
88.7 &
79.1 &
94.8 &
125.0 &
115.9 &
109.9 &
145.1 
\\
Fluc$^\ddagger$ &
37.0 &
25.2 &
20.1 &
26.4 &
23.0 &
20.9 &
22.1 &
20.2 &
17.6 &
18.3 &
18.8 &
14.2 &
\\ \hline
\multicolumn{13}{l}{
$\ast$: Radius from NGC4839, 
$\dagger$: Observational area [arcmin$^2$], $\ddagger$: CXB fluctuation assuming
$S_c=1\times 10^{-13} ~\rm erg~s^{-1}~cm^{-2}$ [\%].
}
\end{tabular}
\label{tab:cxb}
\end{table*}%

\subsection{Results}\label{sec:result}
Temperature and abundance are the basic parameters that 
we derive from the spectral analysis of the ICM.
In order to investigate the radial trend of the temperature and metal abundance,
we analyzed the X-ray spectra in each annuli as described in Sec.~\ref{sec:data}
and Galactic absorption with $N_{\rm H}=1.0 \times10^{20}$ cm$^{-2}$~\citep{dickey90}.
The spectra and the best-fit models for all the annuli are shown in Fig.~\ref{fig:fit}.  
The parameters and the resultant $\chi^2$ values are listed in Table~\ref{tab:bestfit}.
The radial profiles of temperature and abundance are shown in Fig.~\ref{fig:radial}. 
The temperature and abundance within $\timeform{3'}$ from NGC4839 are $kT = 4.18 \pm 0.17$ keV and $Z=0.60\pm0.10~Z_{\odot}$,
respectively, 
which are fairly consistent with 
the XMM-Newton results of $kT = 4.4_{-0.4}^{+0.3}$ keV 
and $Z = 0.36^{+0.14}_{-0.10}~Z_{\odot}$~\citep{neumann01}.
The temperature profile is almost flat at $kT\sim3.6$ keV between $r=10'$ and $r=30'$,
which corresponds to 500~kpc in projection.
These values are also in good agreement with previous measurements~\citep{neumann01, sato11}.
We also confirmed the temperature drop from 3.5 keV to 1.5 keV across the radio relic region.

The abundance profile shows a peak at NGC4839 and gradually decreases with radius from 0.6 to 0.17 solar.
In contrast, the ICM to the north of NGC4839 (gray dotted annuli in Fig.~\ref{fig:image})
has different properties compared with regions at similar distances to the southwest.
The ICM temperature is higher than beyond NGC4839,
as has already been reported by \citet{neumann01}.
They interpreted this temperature structure as a bow shock associated with the infall of NGC4839.
The abundance is lower than to the SW, which may indicate that the SW gas is affected by metal enrichment from NGC4839 via ram-pressure stripping.
We will discuss the enrichment process in the cluster outskirts in Sec.~\ref{sec:abundance}.
In addition, there is a small deviation in the value of the abundance from previous work~\citep{sato11}.
The deviation is possibly explained by the difference in the extraction areas used for the spectral analysis. 

Previous X-ray observations did not find any evidence of a shock associated with the radio relic
in the Coma cluster~\citep{neumann01, feretti06}.
The jump in our temperature profile suggests that there is a shock front in the relic region.
The properties of the shock front will be discussed further in Sec.~\ref{sec:mach}.

The shock front is expected to be located at the outer edge of the relic.
However, in our analysis, the temperature profile appears to break nearer to the inner edge of the relic.  
Previous studies of radio relic clusters (A3667, A3376:~\cite{akamatsu12_a3667, akamatsu12_a3376}) showed
possible evidence of multi-temperature plasma around their relics.
It is likely that this results from image blurring due to the limited PSF of the Suzaku XRT and possibly also projection effects. 
To examine such effects in the radio relic region (\timeform{34'}-\timeform{37'}), 
we fit the spectrum with a two-temperature model.  
We obtained an almost acceptable fit with $\chi^2$/ d.o.f.\ = 204/172
compared with the single temperature case of 208/174.
The resultant hot component temperature was derived to be 
$kT_{high} = 3.56_{-0.54}^{+0.94}$ keV, and the cool one to be 
$kT_{low}$=1.40$_{-0.20}^{+0.56}$ keV while the common abundance was fixed to 0.15 solar 
(two orange crosses in Fig.~\ref{fig:radial}).
These values are almost consistent with those from 
single temperature fits of regions inside ($\timeform{30'}-\timeform{34'}: kT=3.57_{-0.20}^{+0.20}$) 
and outside ($\timeform{43'}-\timeform{54'}: kT=1.54_{-0.39}^{+0.45} $) of the radio relic.

\subsection{Systematic Errors in the Radial Spectral Analysis}\label{sec:sys}
In order to discuss the ICM properties in the cluster outskirts, 
systematic errors must be carefully addressed. 
We considered three components for the systematic error: the NXB estimation uncertainty, 
fluctuations in the CXB intensity, and uncertainties in the contamination on the optical blocking filter.
The NXB uncertainty was estimated by \citet{tawa08} as $\pm$3\%. 
To estimate the fluctuation of the CXB, we follow the method used in~\citet{hoshino10}.
They scaled the fluctuation measured by Ginga~\citep{hayashida89} with the field of view and the flux detection limit.
We removed point sources identified 
by the eye 
in the Suzaku images as shown by the green circles in Fig.~\ref{fig:image}.
We also evaluated the flux of each point source, whose 
lowest value is $8.2\times10^{-14} ~\rm erg~s^{-1}~cm^{-2}$.
To estimate fluctuations in each annular region, 
we adopt a conservative flux limit, $S_c=1\times 10^{-13} ~\rm erg~s^{-1}~cm^{-2}$.
This flux limit is a typical value for the Suzaku satellite \citep{miller12}.
Table~\ref{tab:cxb} shows the resultant CXB fluctuations.
Thus, the estimated CXB fluctuations are in the range of 15-30 \%.

Another potentially large systematic error component
comes from contamination on the blocking filter of the XIS instrument.
After the launch of Suzaku in 2005, the contamination kept increasing for two years, after which it became stable\footnote{http://www.astro.isas.jaxa.jp/suzaku/doc/suzaku\_td/node10.html\#contami}.
All observations analyzed here were performed after 2007.
The amount of contamination is now known with fairly high accuracy,
for which we assign a 10\% error. 
In addition to the above systematic errors,
we also examined the consequences of adopting a different reference
for the solar abundance, namely \citet{anders89} instead of \citet{lodders03}.

The resultant range of best fit parameters after accounting for the systematic errors
are shown in Fig.~\ref{fig:radial} as dotted lines 
in blue for CXB+NXB, green for filter contamination and magenta for different solar abundances.
The effects of systematic errors become larger toward the outskirts.
For most regions, the systematic errors are smaller or comparable to the statistical errors.
In the outermost region, the resulting systematic uncertainty in the ICM temperature is about 0.5 keV.
In the abundance profile, the difference between the two abundance tables is significant around NGC4839.
The difference in the abundances fit with the \citet{lodders03} and \citet{anders89}  models is close to the difference in the iron abundance 
between the two tables.  Thus, the fitted abundances are very similar in absolute terms.

\section{Discussion}
\label{sec:discussion}

\begin{figure*}[t]
\begin{tabular}{cc}
\begin{minipage}{0.5\hsize}
(a) Temperature
\\[-0.8cm]
\begin{center}
\includegraphics[width=.9\hsize,angle=-90]{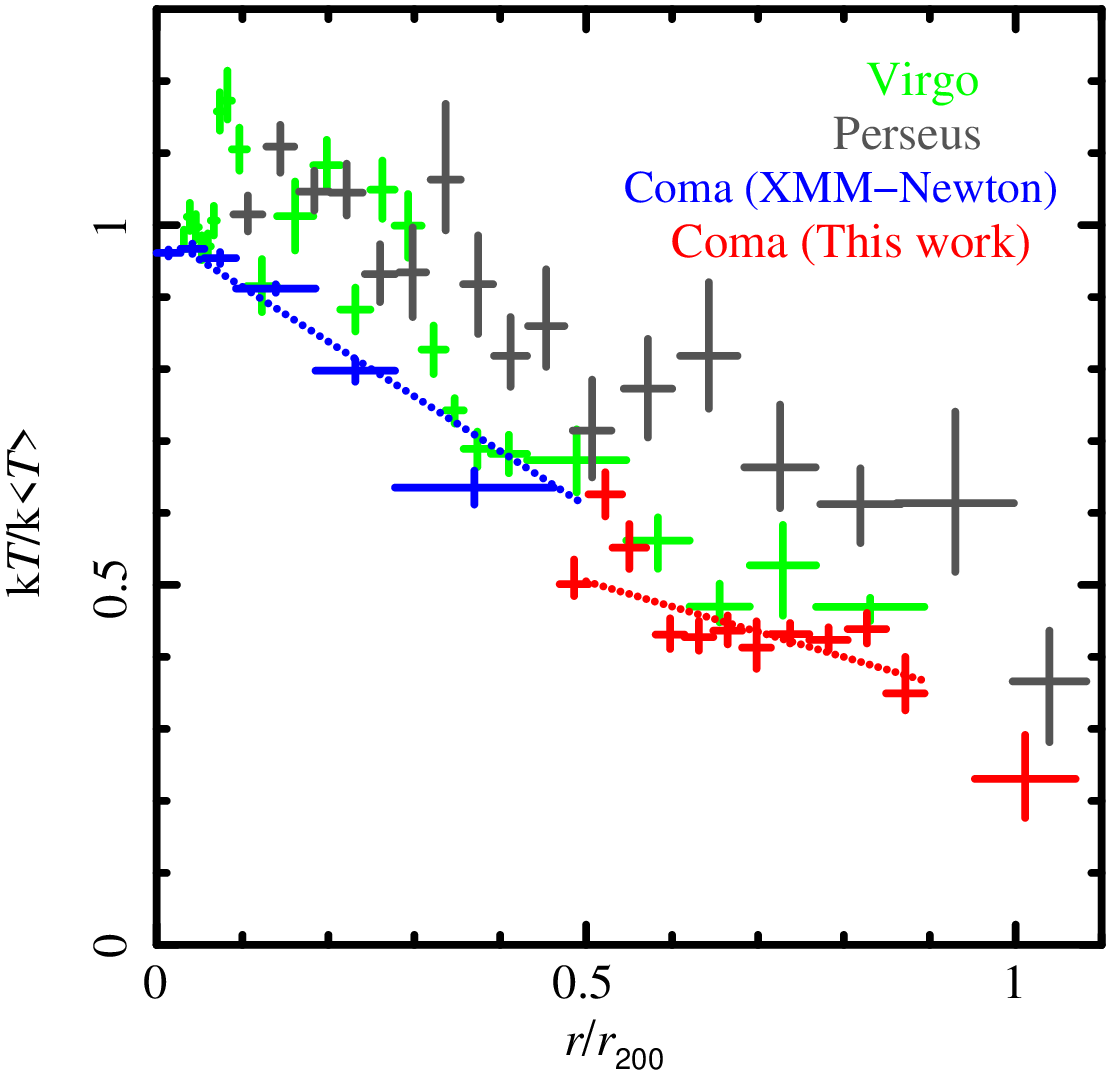}
\end{center}
\end{minipage}
\begin{minipage}{0.5\hsize}
(b) Metal abundance
\\[-0.8cm]
\begin{center}
\includegraphics[width=.9\hsize,angle=-90]{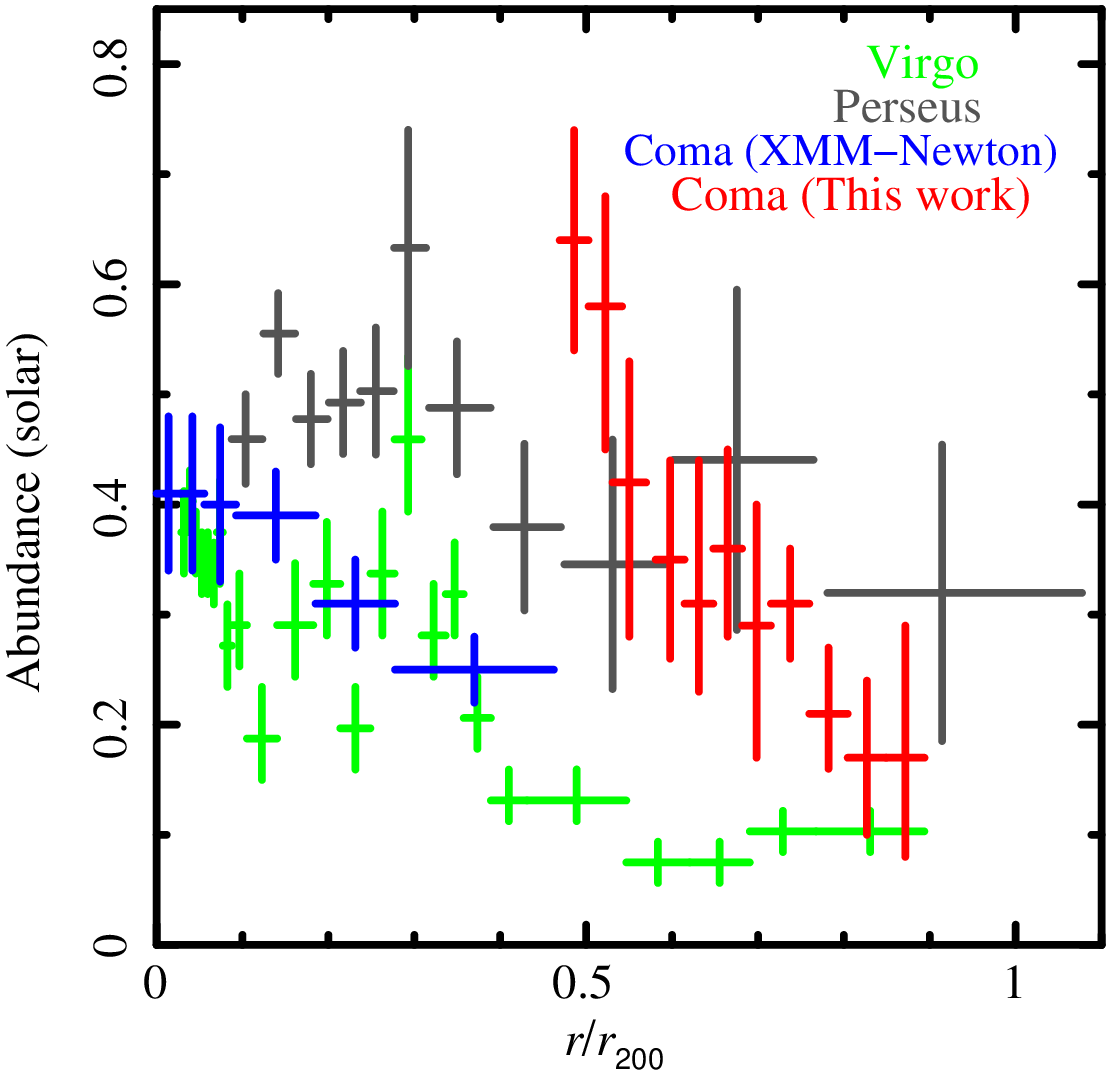}
\end{center}
\end{minipage}\\
\end{tabular}
\caption{ 
Scaled radial profiles of (a) the normalized ICM temperature,  and (b) the metal abundance.
The Suzaku best fit values with statistical errors are shown with red crosses.  
Green and gray crosses show the data from the Virgo and Perseus clusters~\citep{urban11, simionescu11},
and the blue crosses indicate the Coma results from XMM-Newton~\citep{matsushita11}.
To normalize temperature profiles, we adopted average temperatures of 
7.8, 2.3 and 6.4 keV for the Coma, Virgo and Perseus clusters, respectively. 
We estimate $r_{200}$  from the average temperature (see Sec.\ 1).
In (a),  the blue and red dotted line show the best fit linear functions for the XMM-Newton and Suzaku data, respectively.  (See the text for the details.)
In (b), all of the abundances are converted to the~\citet{lodders03} solar values.
 }
\label{fig:combined}
\end{figure*}

Suzaku performed five observations with different pointings of the southwestern regions of the Coma cluster.
The radial profile of the temperature was measured beyond the radio relic region 
as well as beyond $r_{200}$ ($\sim \timeform{83.9'}$ from the center) for the first time.
The ICM temperature declines from $\sim$5.0 keV around NGC4839 to $\sim$3.6 keV at the radio relic.
Furthermore, we find a steep drop in temperature from 3.6 to 1.5 keV beyond the radio relic,
which suggests the presence of a shock front.
The metal abundance also shows a high value of $Z \sim 0.6$ solar at NGC4839 and 
gradually decreases to 0.15 solar around the relic.
Below, we discuss the ICM properties and their implications.

\subsection{Temperature Structure}\label{sec:temp}
The radial profile of the ICM temperature provides information about 
the thermal and merger history of the cluster.
Previous studies of large cluster samples with ASCA and BeppoSAX have indicated
that temperature declines with radius out to at least $\sim 0.4~r_{200}$~\citep{markevitch98, degrandi02}.
These results were extended to 0.7 $r_{200}$ by more recent observations with Chandra and XMM-Newton~\citep{vikhlinin05,pratt07}. 
Suzaku can extend temperature profiles up to $r_{200}$ because of its
 low and stable instrumental
 background~\citep{george08,reiprich09,bautz09,hoshino10,simionescu11,sato12,walker12}.
Previous results indicate that the temperature in relaxed clusters 
shows a systematic drop by a factor of 2-4 from near the center to $r_{200}$.

Here, we discuss the temperature structure in Coma, including
previous XMM-Newton~\citep{matsushita11} and Suzaku~\citep{sato11} measurements.
The profile is shown in Figure~\ref{fig:combined}(a).
The temperature was measured out to $r_{200}$ ($\timeform{83.9'}$), and shows a drop by a factor of 4 
from the center to $r_{200}$. 
The amplitude of this drop is consistent with other clusters.
To quantify the temperature drop, we fit these temperature profiles with a simple linear model,
\begin{equation}
T(r) = \langle T \rangle  \left( a + b \frac{r}{r_{200}} \right) \, ,
\label{eq:linear}
\end{equation}
where we take $k \langle T \rangle = 7.8$ keV as discussed in Sec.\ 1.
We also consider the polytropic function
\begin{equation}
T(r) = T_0  \left[ 1 + \left( \frac{r}{r_c} \right)^2  \right]^{-1.5\beta({\gamma}-1)} \, ,
\label{eq:poly}
\end{equation}
where $T_0$ is the fitted central temperature of the cluster.
The polytropic model was used to fit a large cluster sample in~\citet{markevitch98}.
Previous studies with ASCA~\citep{markevitch98} reported $\gamma=1.24_{-0.11}^{+0.08}$ (90\% error) in 30 clusters.
{
We adopt $\beta$ model parameters for the gas density profile of $\beta=0.72$ and $r_c=11'.1$~\citep{neumann03}.
}
We fit the temperature profile in the inner ($0.1-0.5~r_{200}$) and outer ($0.5-0.9~r_{200}$)  regions 
with these two  models. The resulting values are summarized in Table~\ref{tab:temp_fit}.

\begin{table}[t]
\footnotesize
\caption{Best-fit parameters for the temperature profiles}
\centering
\begin{tabular}{lccc} \hline
	 			&\multicolumn{2}{c}{Linear}    & Polytropic\\
				&$a$	&$b$				& $\gamma$ \\ \hline
\citet{markevitch98} &	-- 	&--	&$1.24_{-0.11}^{+0.08}$\\ 
\citet{pratt07}		&1.19		&$-0.74$	&	--\\ 
This work  ($0.0-0.5~r_{200}$)&$0.99\pm 0.04$ &$-0.76 \pm 0.03$	&$1.23 \pm 0.02^{\ast}$	\\ 
This work  ($0.5-0.9~r_{200}$)	&$0.68 \pm 0.06$ &$-0.35 \pm 0.07$	&$1.29 \pm 0.06^{\dagger}$	\\ \hline
\multicolumn{4}{l}{$\ast$:Fitting range is $0.1-0.5~r_{200}$.}\\
\multicolumn{4}{l}{$\dagger$: $T_0$ set to same value determined in inner region.}
\end{tabular}
\label{tab:temp_fit}
\end{table}%

There is a clear difference in the temperature gradient between the inner and outer regions.
For the inner region,  our best-fit values of the gradient $b = -0.76\pm 0.03$ 
and parameter  $\gamma$=$1.23\pm0.02$ almost agree with previous studies
~\citep{markevitch98,finoguenov01, pratt07}.
On the other hand, in the outer region 
the temperature gradient $b$ is shallower ($b=-0.35 \pm 0.07$).
Because the polytropic model is more sensitive to the inner structure,
it is not surprising that the $\gamma$ parameter does not show much difference 
($\gamma$=$1.29\pm0.06$).
The flatter temperature profile beyond 0.5 $r_{200}$ indicates
 that the outer gas has undergone some additional heating. 
The effect of heating related to merger activity on the ICM temperature is confirmed in many systems~\citep{markevitch07}.
Although we studied a very limited area of the Coma cluster, 
it is likely that the southwest of Coma has experienced heating 
associated with the infall of the NGC4839 group.
In order to explore the effects of more continuous accretion, 
we need temperature profiles in directions which are not affected by recent merger activity.

\subsection{Abundance Structure}\label{sec:abundance}
Measurements of the abundances in the outskirts of clusters are important to study the origin of metals in the ICM. 
Recently, metal abundances at large radii were reported in A399 and A401, and in the Perseus, Virgo and Hydra clusters~\citep{fujita08, simionescu11, urban11, sato12}.
These abundance profiles extend out to $r_{200}$, where the metallicity was found to be 0.1-0.3 solar.
Several metal enrichment mechanisms have been proposed~(e.g., \cite{schindler08}).
One is ram-pressure stripping; the metal-enriched gas in galaxies is stripped by ram pressure
when they move through the ICM at sufficiently high velocities~\citep{gunn72, fujita99}.
HI observations have provided many examples of galaxies being stripped by ram-pressure.
In particular, the Virgo cluster has several spiral galaxies affected by ram-pressure stripping 
(e.g. NGC4522: \cite{kenney04}; NGC4402: \cite{crowl05}; NGC4472: \cite{kraft11}).
X-ray observations by \citet{sun10} found hot tails associated with galaxies {falling into} A3627
(ESO 137-001 and  ESO 137-002).
{Such} stripping events are important not only for galaxy evolution but also for the abundance evolution of the ICM.

\citet{matsushita11} reported radial profiles of the Fe abundance out to 0.3-0.5 $r_{200}$ 
for 28 clusters of galaxies observed with XMM-Newton. 
These clusters show similar abundances within $0.3~r_{200}$ of about 0.4-0.5 solar,
and a decrease to  $\sim$0.3 solar at 0.5 $r_{200}$, using the \citet{lodders03} abundance table.

In the Coma cluster,  our abundance profile shows a significant peak at $0.5 r_{200}$ which corresponds to the position of NGC4839.
Also, abundances outside of NGC4839 ($r >0.5~r_{200}$) are higher than or 
comparable to the central region ($r <0.2~{r_{200}}$).
At the cluster outskirts, 
the contributions from the central region (e.g., the cD galaxy) are {expected to be} small.
In addition, abundances to the SW of NGC4839 (red crosses in Fig.~\ref{fig:radial}) 
are higher than those to the north (grey crosses in Fig.~\ref{fig:radial}).
Since NGC4839 is {most probably moving} 
relative to the ICM from the SW {toward} the NE, 
{facing} an increasing density of gas {in front of it}, 
it is possible that the gas to the SW has been stripped from this galaxy or its associated group.
Our data may provide direct observational evidence that the ICM is enriched by infalling galaxies and groups
at large distances from the cluster center (0.5-0.8 $r_{200}$).

Aside from the high abundance region associated with NGC4839, our data gives the abundance up to 0.9 $r_{200}$.
The observed abundance at large radii decreases to $0.17 \pm 0.11$ solar. 
{This} is similar to previously obtained abundance values at large radii in other clusters 
as mentioned above (Perseus and A399 \& A401).
Recently, a low abundance value was found  at large radii in the Virgo cluster~\citep{urban11}.
They interpreted the low abundance as arising from biasing by the Fe L lines used to determine the abundance.
The Virgo cluster is a low temperature system ($k\langle T \rangle \sim 2.5$ keV) and the temperature at $r_{200}$ has decreased to $\sim$1 keV.
At {such} temperatures, the metallicity {is measured}
primarily from the Fe-L lines,
which are sensitive to the temperature structure~\citep{buote00}.
To examine the effect of the Fe L complex in our Coma spectral fits,
we fitted the $30'$-$34'$ and $34'$-$38'$ annuli in the energy range 1.5-8.0 keV for all the detectors.
The temperature and normalization for each region were allowed to vary, but the abundance was required to be
the same in both regions. 
The resultant ICM abundance was $Z=0.23\pm0.09~Z_{\odot}$, which is consistent with our previous value
($Z = 0.17\pm0.11~Z_{\odot}$).
For the 1.5-8.0 keV energy range, the abundance is determined primarily from the Fe K lines.
Thus, the abundance around $0.9 r_{200}$ in the Coma cluster is likely around 0.2 solar, and is not strongly affected by biasing from Fe L lines.

We confirm the existence of metals in the cluster periphery in Coma along the direction of the merger with the NGC4839 group.
What does this tell us about the origin of these heavy elements?
First, the metals might be due to ram pressure stripping from the NGC4839 group.
When the ram pressure $P_{\rm ram}$ exerted by the external ICM due to the motion
of the NGC 4839 group exceeds the gravitational binding energy density $\epsilon_{\rm grav}$ of gas
internal to the group, the latter can be stripped. Here $P_{\rm ram}=\rho_{\rm ICM} v^2$ 
where $\rho_{\rm ICM}$ is the density of the Coma ICM and $v$ is the infall velocity of NGC4839,
while $\epsilon_{\rm grav}=4GM(r)\rho_{\rm gr}(r)/3r$ where $M(r)$ and $\rho_{\rm gr}(r)$ are the enclosed mass
and gas density of NGC 4839 at radius $r$ from its center. At the current location of
NGC4839, 0.5$r_{200}$ from the Coma center, we can estimate the ICM electron
density before the group's infall to be $\sim 1.5\times10^{-4}\rm ~cm^{-3}$, utilizing the universal
density profile reported by \citet{eckert12} from ROSAT data. Combining this
with $v$ = 1700 km/s as indicated by optical observations \citep{colless96},
the expected value of $P_{\rm ram} \sim 0.6\times10^{-11}$ dyn~cm$^{-2}$.
For the NGC 4839 group, we take a representative radius $r =300$~kpc, corresponding
to $\sim 0.3 r_{200}$ for the group if its virial temperature is 2 keV~\citep{neumann01}
and $r_{200}$ is evaluated as in \citet{henry09}. Assuming a total mass $M\sim5\times10^{13}
M_{\odot}$ \citep{colless96}, we estimate the electron density at $r$=300 kpc before
the infall to be $\sim 4\times10^{-4}~\rm cm^{-3}$, which gives $\epsilon_{\rm grav}(r=300~\rm kpc) \sim 0.6\times10^{-11}~ dyn~cm^{-2}$.
Although subject to many assumptions, this shows that ram-pressure stripping may be marginally possible down to this radius.
After the metals are stripped from the NGC4839 group, they could diffuse due to the turbulence associated with the merger.

Due to projection, the NGC 4839 group might be at a larger radius where the density is lower.
The outer regions of clusters might also be polluted with metals due to galactic superwinds, which may have operated at $z \sim3$ (\cite{ferrara05}) before cluster formation.
Some numerical simulations  \citep{cen06} suggest that 
galactic superwind feedback at high redshifts driven by active galactic nuclei and starbursts 
could provide metals to the proto-cluster gas.

\subsection{Possible Shock Front Associated with the Radio Relic}\label{sec:mach}
\begin{table}[t]
\caption{Shock Properties}
\centering
\small
\begin{tabular}{ccccc}\hline
$T_2$ & $T_1$ & Mach No. & $v_s$ & Compression\\
(keV) & (keV) & ${\cal M}$ & (km s$^{-1}$)  & $C$ \\ \hline
$3.6 \pm 0.2$ & $1.5 \pm 0.4$ & $2.2 \pm 0.5$ & $1410 \pm 110$ & $2.5 \pm 0.4$\\ \hline
\end{tabular}
\label{tab:mach}
\end{table}%

The Coma radio relic, first discovered by \citet{giovannini85}, is one of the best-known relics.
Its location corresponds to a projected distance of $\sim$2.1 Mpc from the cluster center.
The relic measures roughly $800 \times 400$ kpc in the tangential and radial directions, respectively.
\citet{thierbach03} reported a spatially integrated spectral index of $\alpha=1.18\pm0.02$
between 151 MHz and 4.75 GHz, consistent with previous measurements~\citep{giovannini91}.

Our Suzaku data shows a relatively steep temperature drop from inside the relic to outside it towards the southwest.
This suggests that a shock is located somewhere between the middle of the relic and the region just to the
southwest; unfortunately, the limited spatial resolution of Suzaku
does not provide stronger constraints on the location of the shock front.
Shock fronts at the outer edges of radio relics have now been seen in a number of other merging clusters
(e.g., A3667; \cite{finoguenov10, akamatsu12_a3667}).
Since the temperature is high everywhere to the NE of the relic, this indicates that the shock is propagating
towards the SW relative to the pre-shock gas.
Some of the estimated properties of the shock are listed in Table~\ref{tab:mach}.
We estimate the Mach number ${\cal M}$ from the Rankine-Hugoniot condition for the temperature jump,
\begin{equation}
\frac{T_2}{T_1} = \frac{5{\cal M}^4+14{\cal M}^2-3}{16{\cal M}^2} \, ,
\label{eq:tjump}
\end{equation}
where subscripts 1 and 2 denote pre-shock and post-shock values, respectively, and
the assumed ratio of specific heats is $\gamma \equiv 5/3$.
We estimate the post- and pre-shock temperatures based on the temperatures in the annuli from
\timeform{30'}-\timeform{34'} and  \timeform{43'}-\timeform{54'}, respectively.
This gives ${\cal M} = 2.2 \pm 0.5$,
which is similar to the values estimated for other shocks associated with radio relics
(e.g., A3667:\cite{finoguenov10, akamatsu12_a3667}).
With these values, the shock speed is $v_s = (1410 \pm 110)$ km s$^{-1}$.

\begin{figure}[h]
\begin{center}
 \includegraphics[height=1.\hsize,angle=-90]{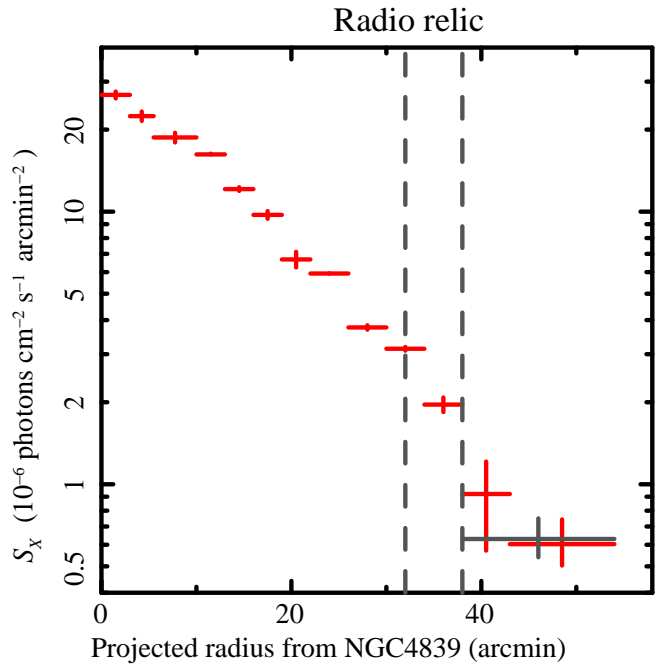}
\end{center}
\caption{Unabsorbed X-ray surface brightness (0.5--10 keV) from the X-ray spectroscopy of the
annuli to the SW of NGC4839.
The red crosses are the same annuli used for the spectra, plus an extra annulus just beyond the
relic which has too few counts for a temperature determination.
The grey cross combines the two annuli beyond the relic to improve the signal to noise.
The dashed vertical lines show the approximate boundaries of the radio relic.
There is a drop in the surface brightness near the outer boundary of the radio relic. }
\label{fig:sb}
\end{figure}

With this value for the Mach number, the shock compression is predicted to be
$C \equiv \rho_2 / \rho_1 = 2.5 \pm 0.4$, where $\rho$ is the gas density.
One can also estimate the shock compression from the jump in the X-ray surface brightness across the relic.
The surface brightness was derived from the observed spectra.
Figure~\ref{fig:sb} shows the X-ray surface brightness (0.5--10 keV) for the same regions used to derived the spectra.
There is a drop in the surface brightness by a factor of $\sim$5.2 near the outer edge of the relic.
Projection effects and the low spatial resolution of Suzaku are likely to cause this to be a slight underestimate of the
surface brightness jump.
For a fixed line of sight path through the gas, this jump in surface brightness corresponds to a shock compression of
$C = 2.1$, including the difference in emissivity for the two regions.
Given that this is probably a slight underestimate, the surface brightness jump is completely consistent with the shock compression of $C = 2.5 \pm 0.4$ predicted based on the temperature jump.

Radio relics generally have rather steep radio spectra, which are believed to be result of radiative losses by the
relativistic electrons which produce the synchrotron emission.
In a number of cases, the radio spectrum is relatively flat at the outer edge of the relic at the location of the shock, and steepens along the direction of flow away from the shock
(e.g., A3667: \cite{roettgering97}, \cite{finoguenov10}; CIZA2242: \cite{vanweeren10}).
The width of the relic and the distance over which the spectrum steepens are consistent with the predicted radiative loss rate and the flow velocity in Abell 3667.
Assuming the strength of the magnetic field is 1-5~$\mu$G, 
the radiative loss time of electrons which radiate at 150 MHz would be $t_{\rm rad} \approx 0.1-0.3$ Gyr (see Fig.~1.6 in \cite{sarazin02});
it is shorter for higher frequency radio emission.
The velocity of the post-shock gas relative to the shock is given by $v_s / C \approx 560$ km s$^{-1}$.
The width of the relic is about 200-400 kpc, and it takes about $t_{\rm ad} \sim 0.5$ Gyr for ICM to travel this distance away from the shock.
Thus, the radiative loss time is likely to be shorter than the advection time for the electrons which emit at all radio frequencies of at least 150 MHz.

If the relativistic electrons are accelerated from thermal energies by diffusive shock acceleration (DSA) and there is no dynamical effect of the relativistic particles or magnetic field on the shock, then the energy spectrum of the relativistic electrons depends only on the shock compression, and is given by $n(E) \, d E \propto E^{-p} \, d E$, where the power-law index is $p = ( C + 2 ) / ( C - 1)$.
For our shock compression, this implies $p \approx 3.0 \pm 0.5$.
If the magnetic field in the relic is roughly constant, such a power-law energy spectrum will lead to synchrotron emission with a spectrum $S_\nu \propto \nu^{-\alpha}$, where
$\alpha = ( p - 1 ) / 2 = 1.00 \pm 0.25$.
This should apply to the radio spectrum immediately behind the shock.
Unfortunately, as far as we are aware, the most recent radio spectral index maps for the Coma radio relic were given by \citet{giovannini91}, and they do not show any clear systematic spectral variations across the relic.
Thus, it is most useful to compare to the integrated spectral index for the relic.
As noted above, we expect radiative losses to be quite effective in the Coma relic.
If the particles lose energy quickly and there is a steady-state between acceleration and radiative losses, then the
particle spectrum steepens by unity,
$p_{\rm loss} = p + 1 \approx 4.0 \pm 0.5$, and the integrated radio spectrum is steeper by one half,
$\alpha_{\rm loss} = \alpha + 0.5 \approx 1.5 \pm 0.25$.
This is steeper than the observed integrated radio spectrum,
$\alpha=1.18\pm0.02$, although the errors in the predicted spectral index are large.
For test-particle DSA, the observed radio spectrum would require a shock compression of $C = 3.21 \pm 0.06$, which is considerably higher than what is suggested by the temperature or surface brightness jump in the shock.
Obviously, it would be useful to have a better spectral index map for the Coma relic, which might allow us to directly determine the spectrum of the most recently accelerated electrons.

The magnetic field strengths assumed above can be justified from upper limits to the non-thermal,
inverse Compton (IC) X-ray emission emanating from the same population of electrons that generate the radio relic.
Relativistic electrons that scatter photons of the cosmic microwave background up to energies 1-10 keV would
have Lorentz factors $\gamma \sim (1-3) \times 10^3$ and radiative loss times $t_{\rm rad} \sim 1-3$ Gyr,
longer than the aforementioned advection time. Hence the electron spectrum at these energies should remain
the same as that at acceleration, and the expected photon index of the IC X-ray emission is $\Gamma=1.68$.
Thus we analyze the XIS spectrum of regions around the radio relic by including a power-law component with
$\Gamma=1.68$ in addition to the thermal ICM emission;
${ apec_{\rm LHB} + wabs (apec_{\rm MWH} + powerlaw_{\rm CXB} + apec_{\rm ICM} + powerlaw_{\rm Nonthermal})}$.
In this paper, 
We account for systematic errors due to contamination on the detector filter
as well as CXB+NXB fluctuations.
For the spectrum of the 34'-38' annular region along the outer edge of the relic (Fig. 3(k)), 
the upper limits on the flux of such a nonthermal component is 
$8.4 \times 10^{-15} ~{\rm erg /cm^2 /s /arcmin^2}$ (0.3-10 keV).
Considering the entire radio-emitting region with area $S = 800 \times 400 {~\rm kpc} \sim 380 ~{\rm arcmin^2}$,
we obtain $3.2 \times 10^{-12} \rm~ erg /cm^2 /s$ (0.3-10 keV).
While the observed radio flux constrains the combination of the magnetic field strength and the density
of non-thermal electrons, upper limits to the IC flux gives an upper limit to the latter alone,
allowing us to derive a lower bound on the field strength of $B > 0.33~ \mu$G.
This can be compared with the result of $B > ~1.05 ~\mu$G based on XMM-Newton data \citet{feretti06}.
It is also in line with values of $B \sim 0.5 ~\mu$G obtained from Faraday rotation measurements
for regions at 2 Mpc distance from the cluster center~\citep{bonafede10}.
We note that our flux limits are also useful for constraining other potential non-thermal emission components
(e.g. \cite{sarazin00, inoue05, vannoni11}).

Although our observations confirm the presence of a shock associated with the Coma radio relic, and the width of the relic is consistent with the flow velocity and radiative loss time scale behind the shock, and the radio polarization implies that the magnetic field is roughly parallel to the long axis of the relic \citep{thierbach03} as expected for a shock,
the observed radio spectrum appears to be flatter than what would be the case for the simplest test-particle DSA theory.
Although the errors in the shock properties we derived are large,
we note that similar results have been found for several other relics, including
the NW relic in Abell~3667 \citep{finoguenov10}, 
and CIZA~2242 \citep{vanweeren10, ogrean12,akamatsu12}.

However, compared to the simplest, test-particle theory,
the physics of DSA under conditions that are realistic and appropriate for clusters 
{can} be more complicated, with no trivial correspondence between the shock Mach number
and the spectral index of accelerated particles.
There are at least two possible ways in which the actual spectrum of accelerated particles can be harder than
the naive expectation from test particle theory, as is observed to be the case here for the Coma radio relic.

The first possibility concerns nonlinear DSA in cosmic ray modified shocks
(see e.g.\ \cite{malkov01} for a review).
If the efficiency of injection into the DSA process is sufficiently high,
the pressure due to the accelerated particles can exert a nonlinear, dynamical back reaction onto the structure of the shock.
The particle spectrum in such circumstances is no longer a simple power-law, but attains an overall, concave shape.
The spectrum at lower energies becomes softer than the test-particle case
as the velocity jump perceived by these particles is reduced by the back reaction,
while the spectrum at higher energies becomes harder
because of the increased compressibility of the downstream gas and consequently enhanced velocity jump.
Nevertheless, this is unlikely to provide an explanation for the observed facts here, since
1) shocks with Mach numbers as low as $\sim 2$ are not expected to foster DSA with efficiencies
high enough for nonlinear effects to become important, 
2) for the conditions relevant to shocks in clusters, the harder spectrum is expected to occur
at energies much higher than for the electrons responsible for the radio relics (e.g. \cite{kang05}),
that is, the radio spectrum may actually be softer than the test particle case if nonlinear effects were effective, 
and 
3) the surface brightness profile shows no evidence of extra compression that might characterize nonlinear shocks, 
and instead the measured compression agrees reasonably with (albeit being slightly smaller than) that expected from 
the temperature jump for a pure gas shock.

The second and more plausible possibility is DSA initiated by injection of pre-existing nonthermal particles,
rather than by that of thermal particles in the ICM \citep{kang11,kang12, pinzke13}.
Before being processed by the low Mach-number shock discussed above,
the pre-shock gas may already contain various types of nonthermal particles,
such as those accelerated by accretion shocks with higher Mach numbers that presumably occur at larger radii \citep{ryu03},
those accelerated by turbulence in regions away from shocks \citep{brunetti11},
those generated continuously as secondary electrons and positrons from collisions of accelerated protons
with thermal protons of the ICM \citep{dennison80,miniati01}, or those from past activity of radio or starburst galaxies.
For DSA by low Mach number shocks, the injection of such pre-existing nonthermal particles
can be much more important than that of particles from the thermal distribution
for which the efficiency is expected to be quite low.
In such cases, the spectrum of accelerated particles may be determined 
by that of the pre-existing population rather than the shock Mach number \citep{kang11,kang12}.
\cite{kang12} and \cite{pinzke13} have specifically shown that such a model can reproduce the observed properties
of some known radio relics associated with shocks detected in X-rays.
Although a quantitative discussion is beyond the scope of this paper,
we note that 
unlike in the central regions of the cluster, the ICM density in the cluster outskirts is low enough so 
that nonthermal electrons can survive without significant energy losses over very long times 
(longer than a Hubble time for some energies, see Fig.1 of \cite{pinzke13}). 
The low density also means that injection of secondaries from proton-proton collisions is inefficient, 
so the most likely source of the pre-existing electrons is direct acceleration
by shocks of higher Mach number in the past. 
For the Coma radio relic, they may have originated from past activity of nearby galactic winds or radio galaxies,
perhaps from within the NGC 4839 group before its passage through the virial radius,
the putative accretion shock at larger radii, or even past superwind activity,
after which they were advected inward along the large-scale structure filament from the SW.

\subsection{Origin of the Shock Front at the Radio Relic}

In modern theories of hierarchical structure formation,
both internal merger shocks and external accretion shocks around clusters of galaxies are
predicted \citep{miniati00, ryu03}.
External accretion shocks are located at large radii (about 3 times the virial radius),  are more spherical in shape,
and are expected to have high Mach numbers of up to $10^3$.
Internal merger shocks located around the virial radius have Mach numbers ranging between 2-10 in 
{the} simulations, and are smaller in extent.
The temperature jump associated with the Coma radio relic located at 2 Mpc (0.9 $r_{200}$)
implies a Mach number of ${\cal M} \sim 2$, which is not compatible with expectations for an external accretion shock 
{in} the diffuse intergalactic medium.
This might be an outgoing merger shock associated with the infall of the NGC4839 group.
The tail and abundance structure around NGC4839{, as well as optical observations,} 
suggest that it is moving to the NE, so the relic shock is obviously not  a merger bow shock in front of this group.
Another possibility is that the relic shock is due to accretion~of diffuse gas 
from the large scale structure filament \citep{brown11}
which extends from Coma SW towards the Abell 1367 cluster.
Recently, \citet{kale12} discussed the effect of large-scale accretion flows on the outgoing merger shocks
based on their radio observations of the typical major merger cluster Abell 3376.
Recent numerical simulations \citep{paul11} indicate that shock fronts are enhanced
along filament directions.
In addition, the radio morphology of the radio galaxy NGC4789 suggests that it is affected by an accretion flow onto the Coma cluster from the large scale structure filament
(see Fig.~1 in \cite{giovannini91}).
Considering the lower ICM temperature in the pre-shock region and the evidence for infall of NGC4839, 
the shock may be generated by the combination of 
the merger between Coma and NGC4839 and the inflow of cooler gas
from the large scale structure filament towards Abell~1367.

Our data may provide direct observational evidence of a shock involved with the infall of the NGC4839 group.
Such shocks transform  the kinetic energy of the merging subcluster into the thermal energy of the ICM, and this process 
is fundamental for  the thermal history of clusters of galaxies.
As shown in  our previous work \citep{akamatsu12}, radio relics are good tracers of large scale structure shock fronts.
Further studies (radio, SZ, X-rays, and simulations) are needed in order to further clarify
the nature of these shock fronts.

\section{Summary}
We observed the NGC4839 group and the Coma radio relic with the Suzaku XIS and derived
radial profiles of the temperature and the metal abundance. 
The temperature across the radio relic shows a jump, which indicates the presence of a 
shock front.
We estimated a Mach number of ${\cal M} \approx 2.2$ using
the Rankine-Hugoniot temperature jump condition.
 
The main results of this work are summarized as follows;
\begin{itemize}
\item The ICM temperature distribution is  flat between $10'-30'$ from NGC4839, and then
decreases sharply (consistent with a ${\cal M}\sim2.2$ shock) at some location near the outer edge of the radio relic.
\item Compared to the inner regions, there is a clear change in the temperature structure around NGC4839.
\item The ICM abundance has a peak at NGC4839 and gradually decreases from 0.6 to 0.2 solar towards the
outer regions. 
The abundance profile shows an excess over the mean values for
other clusters at such large radii as observed with XMM-Newton.
\item The estimated Mach number of the shock is ${\cal M}$=2.2$\pm0.5$ based on the temperature jump, which corresponds to a shock velocity of $v_s = (1410 \pm 110)$ km s$^{-1}$.
The X-ray surface brightness also shows a drop near the outer edge of the radio relic.
Applying the shock density jump condition to the surface brightness profile gives a Mach number which
is consistent with the value from the temperature jump.
The implied post-shock velocity and the rate of radiative losses by the radio-emitting relativistic electrons imply
that losses are very significant, and can help to explain the relatively narrow width of the relic.

\item The magnetic field strength of the radio relic region is constrained to be 
$B > 0.33 ~\mu$G from upper limits to the inverse Compton X-ray emission, a bound that is somewhat
  weaker than but consistent with previous similar measurements.

\item The observed integrated radio spectrum of the relic is somewhat flatter than what would be predicted by test-particle diffusive shock acceleration (DSA) in a shock with the observed Mach number and compression.
A similar result is seen in other relics.
We discuss possible explanations for this discrepancy, and conclude that the shock may be re-accelerating an existing population of low energy relativistic electrons, rather than directly accelerating thermal electrons to high energies.
This might also explain why some shocks are seen in clusters without associated radio relics.
\end{itemize}
These results show that the NGC 4839 group is falling into the Coma cluster, and that
the radio relic was possibly generated by shock acceleration.
The origin of the shock structure is still unclear.
Upcoming radio (LOFAR: \cite{vanweeren12}) and SZ (ALMA: \cite{yamada12}) observations will 
provide new insights into the nature of the shock front and the dynamical history of the Coma cluster.
Obviously, additional wide field X-ray observations would also be very useful.

As this work was being completed, we became aware of the preprint by \citet{ogrean12_coma},
who report evidence for a shock with Mach number ${\cal M}\sim 2$ at the Coma radio relic based 
on archival XMM-Newton data. 
More recently, \citet{simionescu13} reported the results of mapping observations of the Coma cluster by Suzaku.

\bigskip 
The authors would like to thank all the members of the Suzaku team for their continuous contributions 
in the maintenance of onboard instruments, calibrations, software development.
We would like to thank Toru Sasaki, Takaya Ohashi and Jelle de Plaa 
for careful reading of the manuscript and useful discussions.
We are grateful to the referee for useful comments that helped to improve this paper.
SI was supported by Grants-in-aid from MEXT of Japan Nos. 22540278 and 24340048.
CLS was supported in part by NASA Suzaku grants NNX09AH74G and NNX08AZ99G, by NASA ADAP grant NNX11AD15G, and by Chandra grant GO1-12169X.



\end{document}